\documentclass[preprint]{aastex}

\shorttitle{TeV emission from Galactic Center}
\shortauthors{Aharonian and Neronov}

\usepackage{graphicx}
\newcommand{\beq}{\begin{equation}}
\newcommand{\eeq}{\end{equation}}
\newcommand{\beqa}{\begin{eqnarray}}
\newcommand{\eeqa}{\end{eqnarray}}

\def\l({\left[}
\def\r){\right]}

\begin{document}


\title{High energy gamma rays from the massive black hole\\ in the galactic center}
\author{F.Aharonian}
\affil{Max-Planck Institut f\"ur Kernphysik,\\ D-60029 Heidelberg, Germany}

\author{A.Neronov}
\affil{Ecole Politechnique Federale de Lausanne, \\ BSP, 1015, Lausanne, Switzerland}

\begin{abstract}
Accreting Black Holes (BHs) are believed to be sites 
of possible  particle acceleration with favorable conditions 
also for  effective gamma-ray production.  However, because 
of photon-photon pair production, only low energy (MeV) 
gamma-rays can escape these compact objects with
typically very large compactness parameter, $\kappa= \frac{L}{L_{\rm
Edd}} \frac{R_{\rm g}}{R} \geq 0.01$, given that in most cases the accretion
disks within 10 Schwarzschild radii $R_{\rm g}$ radiate with a power exceeding 
10 percent  of the Eddington luminosity, $L_{\rm Edd}$.  Therefore the high energy
gamma-ray emission of these objects (both of stellar mass and
super-massive BHs) is generally suppressed, and consequently the
unique information on possible particle  acceleration processes near the 
event horizon of the BH is essentially lost.  Fortunately this
is not the case for the super-massive BH located at  
the dynamical center of our Galaxy (Sgr~A*), which thanks to its extraordinary low
bolometric luminosity ($ \leq 10^{-8} L_{\rm Edd}$) is transparent for
gamma-rays up to very high energies, $E \sim 10 \ \rm TeV$. We discuss
different scenarios of gamma-ray production in Sgr~A*, and show that
for a reasonable set of parameters one can expect detectable gamma-ray
fluxes of both hadronic and electronic origin. Some of these scenarios
are applicable not only for the TeV gamma-ray emission recently
reported from the direction of Galactic Center, but may have broader
implications relevant to highly variable nonthermal emission of Sgr~A*
in radio, IR and X-ray bands.
\end{abstract}

                  
\maketitle

\section{Introduction.}
\label{intrd}

The central 10 pc region of our Galaxy is an extraordinary site that harbors
many interesting sources  packed  with an unusually high density  around the most 
remarkable object of this region,  the  compact radio source Sgr~A$^\ast$.  The latter 
most likely  associates  with a hypothetical super-massive black hole (BH), 
$M \approx 3   \times 10^6 M_\odot$, located very close to  the dynamical center of the 
Galaxy \cite{genzel2000,ghez2000,schoedel2002}.  The upper limits on the size of
the source at mm-wavelengths on the level of $\sim 0.1$
milli-arcseconds (e.g. \cite{krichbaum1998}) tell us that the
emission is produced within 10 Schwarzschild radii around BH.  The
time variability recently detected at X-rays 
\cite{baganoff2001,porquet2003,goldwurm2003} and near-infrared
wavelengths (e.g. \cite{genzel2003})
on $\leq 1 \ \rm h$ timescales is an independent  evidence that the radiation
comes from regions located very close to the event horizon of BH.

The temporal and spectral features of radiation of Sgr A$^\ast$ are
quite unusual and, as a whole,  essentially different from other compact
galactic and extragalactic sources containing BHs.  
This concerns, first of all, the extraordinary low luminosity of  Sgr A$^\ast$.
Many scenarios  have been proposed to explain this effect 
invoking, in particular,  advection   \cite{narayan1995} and convection 
\cite{quataert2000,narayan2002,igumenshchev2002}   dominated accretion flow models, 
advection dominated inflow-outflow solution \cite{blandford1999},
inefficient accretion flow  model \cite{yuan2003}), Bondi-Hoyle
type models \cite{review}, Jet models \cite{falcke2000}, models
assuming interactions of stars with cold accretion disk \cite{nayakshin2003}, {\em etc}.  
Concerning the radiation mechanisms, there is little doubt that the emission components 
at radio, and most likely also at   the  IR and X-ray wavelengths 
have nonthermal origin. The currently most favored models are 
different versions of the so-called Synchrotron-Self-Compton (SSC) scenario which
assumes that the radio and mm emission is due to  
electron synchrotron radiation, while the X-rays are explained by inverse 
Compton scattering of the
same (relatively low energy) electron population (see for a review
\cite{review}).  On the other hand, if the recently detected
TeV emission from the Galactic Center also comes from the inner parts of 
Sgr A$^\ast$, this would imply that relativistic particles (protons and/or electrons), are
accelerated to very high energies in the vicinity of the BH. 
Below we  show that these particles  play  non-negligible role in the 
formation of  energy spectrum of radiation at low frequencies as well.
This allows decisive tests of models of very high energy gamma-radiation through 
simultaneous multiwavelength studies of  Sgr A$^\ast$. 

\subsection{Broad-band observations of Sgr~A*: a short overview}
\label{overview}

\begin{figure}
\begin{center}
\resizebox{0.7\hsize}{!}{\includegraphics[angle=0]{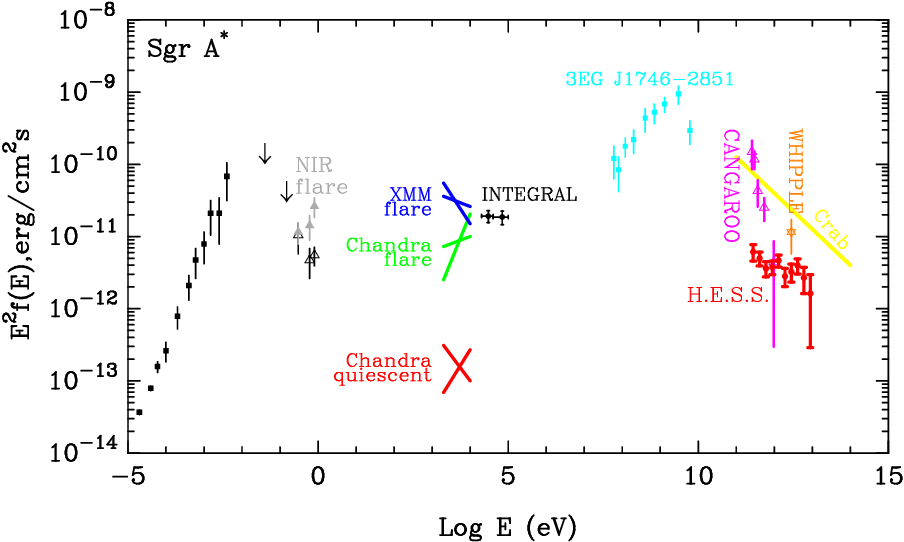}}
\caption{
Broad-band spectral energy distribution of  Sgr A*. Radio data  are
from \cite{zylka1995}, and the  infrared data for quiescent state 
and for flare are from \cite{genzel2003}.  X-ray fluxes measured by Chandra 
in the quiescent state  and during a  flare  are from
\cite{baganoff2001,baganoff2003}.  XMM measurements of the X-ray flux  
in a flaring state  is from \cite{porquet2003}. In the same plot we show also 
the recent  INTEGRAL  detection of a hard X-ray flux, however 
because of relatively poor angular resolution the relevance of this flux to Sgr A* 
hard  \cite{belanger2004} is not yet established.   The same is true also for the 
EGRET data \cite{mayer-hasselwander1998}  which do not allow localization
of the GeV source with accuracy better than 1 degree. 
The very high energy  gamma-ray fluxes   
are obtained by the CANGAROO \cite{tsuchiya2004},
WHIPPLE  \cite{kosack2004} and H.E.S.S. \cite{aharonian2004} groups. 
Note that  the GeV and TeV gamma-ray fluxes  
reported from the direction of the Galactic Center  may originate in sources 
different from  Sgr~A*, therefore, strictly speaking,  they should be considered
as upper limits of radiation  from  Sgr~A*.
}
\label{fig:SGR_data}
\end{center}
\end{figure}

The radio to mm radiation of Sgr~A* is characterized by a very hard
spectrum (see Fig.~\ref{fig:SGR_data}) with spectral index $\alpha\simeq 0.3$ ($F_\nu\sim
\nu^\alpha$), low-frequency turn-over at $\nu\simeq 1$ GHz and
high-frequency cut-off at $\nu\simeq 10^3$ GHz \cite{zylka1995}. 
The hard spectral index can be explained by optically thin 
synchrotron emission from Maxwellian type energy distribution of 
relativistic  electrons \cite{duschl1994,beckert1996}
or by synchrotron self-absorption of radiation in an optically thick source \cite{selfabsorb}. 
It should be noted, however, that the measurements of scattering of photons by interstellar
plasma indicate \cite{lo1998,bower2004}  that the radiation at different wavelengths 
is produced at different distances from BH.  Namely while the mm  
emission originates from a compact region of a  size
$R_{\rm ir}\simeq 20 R_g$ ($R_{\rm g} = 2 G
M/c^2 \simeq 10^{12} \ \rm cm$ is the gravitational radius of
the BH in GC ), the radio emission is produced  at larger distances. 
 On the other hand,  the near-infrared and X-ray flares,  with variability time scales 
$t_{\rm IR} \sim 10^4$ s \cite{genzel2003} and 
$t_{\rm x}\sim 10^2-10^3$ s \cite{baganoff2001,porquet2003},   
indicate that the radiation at  higher frequencies  is produced 
quite close to  the BH horizon. It has been shown recently by \cite{liu+melia}
that acceleration of moderately relativistic electrons ($\gamma_{\rm e} \sim 100$)
by plasma wave turbulence near the BH event horizon and subsequent spatial 
diffusion of highest energy electrons can explain 
the wavelength-dependent size of the source. The same electron population
can explain the X-ray flares through the inverse Compton scattering due to
dramatic changes of physical conditions during the flare \cite{markoffetal2001,liu+melia}

Very hard X-ray emission  up to 100 keV, with a  possible detection of a 40min flare 
from the central 10 arcmin region of the Galaxy recently has  been reported by the 
{\it INTEGRAL} team \cite{belanger2004}.

In the $\gamma$-ray band 100 MeV-10 GeV gamma-rays from
the region of GC have been reported by the EGRET team
\cite{mayer-hasselwander1998}.  The luminosity on MeV/GeV
gamma-rays $L_{\rm MeV/GeV} \simeq 10^{37} \ \rm erg/s$ 
exceed by an order of magnitude the
luminosity of Sgr A* at any other wavelength band, see Fig.~\ref{fig:SGR_data}. 
However, the angular resolution of EGRET was  too large to distinguish between the diffuse
emission from the region of about 300~pc and the point source at
location of Sgr A*.  GLAST, with significantly improved  (compared to EGRET) 
performance can provide higher quality images of this region, as well as  
more sensitive searches for variability of GeV emission.  This would allow 
more conclusive statements  concerning the origin of MeV/GeV gamma-rays.

TeV gamma radiation from the GC region recently has  been reported by
the CANGAROO \cite{tsuchiya2004},  Whipple \cite{kosack2004} and H.E.S.S. 
\cite{aharonian2004}  collaborations.  Amongst possible sites of production of 
TeV signals are the entire diffuse 10 pc region as result of interactions of
cosmic rays with the dense ambient gas, the relatively young supernova
remnant Sgr A East \cite{fatuzzo2003}, the Dark Matter Halo \cite{bergstrom1998,gnedin2004} 
due to annihilation of
super-symmetric particles, and finally Sgr A* itself.  It is quite
possible that some of these potential gamma-ray production sites
contribute comparably  to the observed TeV flux.  
Note that  both the energy spectrum and the flux measured by H.E.S.S. \cite{aharonian2004}
differ significantly from the results reported by CANGAROO  \cite{tsuchiya2004} 
and Whipple \cite{kosack2004}  groups (see Fig.~\ref{fig:SGR_data}). 
If this is not a result of miscalibration  of  
detectors, but rather is due to the variability of the source, Sgr~A* 
seems to be the most likely candidate to which the TeV radiation could be associated,
given the localization of a point-like TeV source by H.E.S.S. within 1 arcmin around Sgr~A*.  
 But for unambiguous conclusions, one needs 
long-term continuous monitoring of  the GC region 
with well calibrated TeV  detectors,   and especially 
multiwavelength   observations  of Sgr~A*   together 
with  radio, IR  and X-ray telescopes.
With the potential to detect short ($\leq$ 1 h) gamma-ray flares  at  the
energy flux level below $10^{-11} \ \rm erg/s$, H.E.S.S.  should be 
able to provide meaningful searches  for variability of TeV
gamma-rays on timescales less than 1 hour which is crucial for
identification of the TeV source with Sgr A*.

In this paper we assume that Sgr~A* does indeed emit TeV gamma-rays,
and explore possible mechanisms of particle acceleration and radiation
which could lead to production of very high gamma-rays in the immediate vicinity
of the associated supermassive black hole.  At the same time,  since the origin of TeV
radiation reported from the direction of GC is not yet established,
any attempt to interpret these data quantitatively  would be  rather premature and inconclusive. 
Moreover, any model calculation of TeV emission of a compact source with characteristic 
dynamical time scales less than 1 hour would require  data obtained at different
wavelengths {\em simultaneously}.  Such data are not yet
available for Sgr~A*. Therefore in this paper we present
calculations  for  a set of generic model parameters with a general  aim to
demonstrate the ability (or inability) of certain models for  production of 
detectable fluxes of TeV gamma-rays without violation  the data 
obtained  at radio, infrared and X-ray bands  (see Fig.~\ref{fig:SGR_data}).  More
specifically, we discuss  the following possible models in which TeV
gamma-rays can be   produced due to (i) synchrotron/curvature radiation of
protons, (ii) photo-meson interactions of highest energy protons with 
photons of the compact IR source, (iii) inelastic p-p interactions
of multi-TeV protons in the accretion disk, (iv)  Compton cooling  of multi-TeV 
electrons  accelerated  by induced electric field  in the vicinity of  the massive BH.    

\section{Internal absorption of $\gamma$-rays}
\label{sec:data}

The very low bolometric luminosity of Sgr A* makes this object rather unique
among the majority of galactic and extragalactic compact objects containing 
black holes. One of the  interesting consequence of the faint  electromagnetic radiation of  
 Sgr~A*   is that  the latter  appears transparent for gamma-rays up to very high energies ! 
Thus the TeV studies of Sgr A* open  unique opportunity to study high energy processes 
of particle acceleration and radiation in the  immediate vicinity of the event horizon. 
In this regard one should note  that TeV gamma-rays observed from several  
BL Lac objects (a subclass of  AGN) originate in relativistic jets quite far from the 
central compact engine, therefore they do not carry direct information  about the 
processes  in the vicinity of the central BH.    

Generally,  the AGN  cores and X-ray binaries harboring  supermassive and stellar mass 
black holes,  are characterized by very dense  ambient photon fields which do  not 
allow the high energy gamma-rays to escape  freely  their  production regions.    
In the isotropic field of background photons, the cross-section of photon-photon pair 
production depends on the product of colliding photons, $s=E \epsilon /m_{\rm e}^2 c^4$.
Starting from the threshold at $s=1$, the cross-section 
$\sigma_{\rm  \gamma \gamma}$ rapidly increases achieving the  maximum 
$\sigma_0 \approx \sigma_{\rm T}/5 \simeq 1.3 \times 10^{-25} \ \rm cm^2$ at 
$s \approx 4$, and then decreases as $s^{-1} {\rm ln} s$. Because of relatively
narrow distribution of $\sigma_{\rm \gamma \gamma}(s)$, gamma-rays interact 
most effectively with the background photons of  energy
\beq
\epsilon_b \approx 1 (E /1 \ \rm TeV)^{-1} \  eV \ .
\label{epsilon}
\eeq
Thus the optical depth for a gamma-ray of energy $E$ in a source of
luminosity $L_\epsilon$ at energy given by Eq.(\ref{epsilon}) and size
$R$ can be written in the form
\beq
\label{opt}
\tau (E) =\frac{L_\epsilon \sigma_{\rm T}(E)}{4 \pi Rc\epsilon_b}\simeq
10^8\left[\frac{L_\epsilon}{L_{\rm Edd}}\right]
\left[\frac{R_{\rm g}}{R}\right] \left[\frac{E}{1\mbox{ TeV}}\right] .
\eeq
Here the optical depth is normalized to the compactness parameter
$\kappa=(L_\epsilon / L_{\rm Edd}) (R /R_{\rm g})^{-1}$  which does not depend on
the mass of BH, therefore  is applicable to both stellar mass and
super-massive BHs.  One can see from this estimate that for a typical
AGN or an  X-ray binary with NIR/optical luminosity 
$L \geq 10^{-5} L_{\rm Edd}$, TeV gamma-rays cannot escape the source unless 
they are produced far from BH, at distances exceeding $10^3$ Schwarzschild radii.

The luminosity of Sgr A* is unusually low for an accreting  massive BH. At
NIR/optical wavelengths the luminosity does not exceed $10^{-8}$,
therefore TeV gamma-rays can escape the source even if they are
produced at $R \sim R_{\rm g}$. 
Numerical calculations of the optical depth based on the spectral
energy distribution of Sgr~A*  shown in  Fig. \ref{fig:opt_depth} 
confirm this conclusion.  It is seen that indeed only at energies above 10 TeV
the absorption of gamma-rays becomes significant, even if one assumes that
the production region of radiation is limited within $2 R_{\rm g}$.

It is interesting to note that the decrease of the pair production
cross-section well above the pair production threshold ($s \gg 1$)
makes  the source  again transparent, but  at EeV
energies.  These gamma-rays are not absorbed on they way to the Earth
either and can be detected  by arrays  of highest energy cosmic
rays like AUGER. However, at such large energies gamma-rays can be absorbed due to 
pair production in the magnetic field inside  the source.

The mean free path of a gamma-ray photon of energy $E$
in a magnetic field of strength $B$ can be approximated as \cite{erber1966}
\begin{equation}
\label{gamma_B}
\Lambda_{B\gamma}\approx \frac{2\hbar E}{0.16\alpha_fm_ec^3K_{1/3}^2(2/3\xi)} \,
\end{equation}  
where  $\xi=(E/m_ec^2)(B/B_{\rm cr})$, $B_{\rm cr}=4.4 \times 10^{13} \ \rm G$
and $K_{1/3}(x)$ is the modified Bessel function.
The mean free path of gamma-rays as function of energy for  different
magnetic fields is shown in Fig.~\ref{fig:lambdaB} \footnote{Note, that $B$ which enters 
in Eq.~(\ref{gamma_B}) through the parameter $\xi$, 
is the component of magnetic field normal to the photon momentum. This means 
that the absorption length of the gamma-ray  propagating along the lines of an
ordered magnetic field can be larger than the mean free paths shown in 
Fig.~\ref{fig:lambdaB}}.  
It is seen that the mean free path of $\geq 10^{17} \ \rm eV$  gamma-rays in the magnetic 
field of strength $B=10  \ \rm G$ becomes shorter  than the gravitational  radius of the black hole 
of mass $3 \times 10^6 M_\odot$. For very strong magnetic fields, $B \geq 10^6  \ \rm G$,  
the source is  opaque for  $\geq 1 \ \rm TeV$  gamma-rays as well.   

Generally,  the  process of interactions of gamma-rays with magnetic field cannot be  
reduced to  a simple   absorption effect. Indeed, the secondary  electrons interacting 
with radiation  and  magnetic fields  produce  new  gamma-rays,  which  in turn 
lead to  a new generation of electrons-positron pairs; thus,  
a nonthermal cascade develops
the features  of which strongly depend on the energy densities of photon and magnetic fields
(see e.g. Aharonian \& Plyasheshnikov 2003). Note  that since  the 
pair production of gamma-rays in the $(B,E)$ parameter space of  interest 
($E \leq 10^{18} \ \rm eV$ and $B \leq 10^6 \ \rm G$) 
always takes place in the regime when  $\xi=(E/m_ec^2)(B/B_{\rm cr}) \ll 1$, interactions  
with  the magnetic field quickly lead  to degradation of the energy of leading particles
(synchrotron photons are produced with energies far below  the energy of 
the parent electrons, therefore cannot support effective development of the cascade).

Generally, Klein-Nishina cascades in photon fields last longer, however in  the presence of 
even relatively week magnetic field they  can be strongly suppressed due to 
synchrotron cooling of electrons.  In Sgr~A* where the energy density of 
low-frequency radiation  is estimated $w_{\rm rad} \sim 1 (R/10 R_{\rm g})^{-2} \ \rm erg/cm^3$, 
for effective development of  an electromagnetic cascade  
the strength of the magnetic field should  not exceed 
$(8 \pi w_{\rm rad})^{1/2} \simeq 5 \ \rm G$.

\section{Gamma Ray  emission mechanisms}

High energy gamma-rays from compact regions close to the event 
horizon of a  massive black can be produced in various ways  due to acceleration of
protons and/or electrons and their interactions with ambient magnetic
and radiation fields, as well as with the  thermal plasma.  Below we discuss 
the basic features of different  possible gamma-ray production scenarios in Sgr A*.

\subsection{Gamma-rays related to accelerated protons}

\subsubsection{Synchrotron and curvature radiation of protons} 
\label{pB}
Gamma-rays produced by relativistic protons and nuclei are often
called  ``hadronic gamma-rays''. However, this is not the case of
interactions of protons with magnetic field. Namely,   the  
mechanisms associated with the synchrotron and curvature radiation components 
have electromagnetic origin. These processes become
important in compact, strongly magnetized  astronomical environments.
Moreover, they  are unavoidable in the so-called
extreme accelerators in which  particles are accelerated at the
maximum possible rate, $\dot{E}=eB$, determined by classical
electrodynamics \cite{aharonian2002}.  In the  case of the 
Galactic Center, protons can be boosted to the
maximum possible energy, assuming that acceleration takes place 
within 10 gravitational radii around the central black hole: 
\begin{equation}
\label{max_p}
E_p \sim eBR \simeq 10^{18} \left[\frac{B}{10^4\mbox{ G}}\right]
\left[\frac{M}{3\times 10^6M_\odot}\right]\mbox{ eV} \ .
\end{equation}

Thus,  for the given magnetic field $B$, the synchrotron radiation
formally could extend up to the characteristic energy
\begin{equation}
\label{max_hnu}
\epsilon  \sim  0.1   \left[\frac{B}{10^4 \mbox{ G}}\right]^2
\left[\frac{M}{3 \times 10^6M_\odot}\right]^2 \mbox{ TeV} \ .
\end{equation} 

However, even for an ``ideal'' combination of parameters allowing the
most favorable acceleration/cooling regime when the proton
acceleration proceeds at the maximum rate and the energy losses are
dominated by synchrotron cooling, the characteristic energy of
synchrotron radiation is limited by         
\begin{equation}
\label{p_synch_cut}
\epsilon_{\rm  max}=\frac{9}{4\alpha_f}m_pc^2\simeq 0.3 \mbox{ TeV} \ ,
\end{equation}
which does not depend on the strength of magnetic field ($\alpha_f=1/137$ is
the fine-structure constant). This leads to the self-regulated
synchrotron cut-off \cite{aharonian2000} at  $\epsilon_{\rm cut}=a
\epsilon_{\rm c, max}$, where the parameter $a$  varies
between 0.3 in the case monoenergetic electrons and $\sim 1$ for
power-law distribution of electrons with an exponential cutoff.

This implies that the proton synchrotron radiation cannot explain the
gamma-ray flux observed from the direction of GC up to several TeV,
unless the radiation takes place in a source  relativistically moving towards 
the observer with bulk  motion  Lorentz  factor exceeding 10.
Another possibility for extension of the high energy end of the
spectrum to multi-TeV domain can be realized if the proton
acceleration and $\gamma$-ray production regions are separated from
each other.  For example, assuming that protons are accelerated in a
regular field configuration while moving along field lines, and later
are injected into a region with strong chaotic magnetic field, we may
avoid the upper given by Eq.(\ref{p_synch_cut}).
Nevertheless it cannot be arbitrary large because even in a regular
filed charged particles suffer radiative losses due to curvature
radiation.  Note that  as long as we are interested in high energy
nonthermal emission  the curvature radiation should not be treated as a 
source of energy losses but rather a radiative process with a 
non-negligible contribution to the gamma-ray emission of the
accelerator. In the case of the black hole in GC this contribution could
be quite significant \cite{levinson2000}.

Compared to synchrotron radiation the spectrum of curvature radiation
of protons can extend  to  higher energies. Assuming that proton
acceleration proceeds at the maximum possible rate, $\dot{E} \sim eB$,
and  is balanced by losses due to curvature radiation, one arrives at
the following estimate of the maximum photon energy
\begin{equation}
\label{gamma_curv}
\epsilon_{\rm  max} ={3 E_p^3\over 2 m^3R}\simeq 0.2  
\left[{B\over 10^4 {\rm G}}\right]^{3/4} \mbox{TeV}
\end{equation} 

Formally, Eq. (\ref{gamma_curv}) allows extension of the spectrum of
curvature radiation to 10 TeV, if the magnetic field exceeds $B \simeq 10^6$ G. 
However,  as discussed in the previous section, 
for such a strong magnetic field, the source is not transparent for  TeV 
$\gamma$-rays   (see  in Fig.~\ref{fig:lambdaB}).

\subsubsection{Photo-meson interactions} 

Protons can produce TeV radiation through interactions with
 ambient photon fields. The photo-meson processes are especially 
effective at energies $\sim 10^{18} \ \rm eV$ 
because such energetic  protons start to interact with the
most copious, far infrared and mm, photons.  Despite the low
luminosity of Sgr A* , $L_{\rm mm} \simeq 10^{36}$ erg/s, because of
the small source size (e.g. Melia and Falke 2001) the density of
infrared photons appears sufficiently high,
\begin{equation}
\label{backgr}
n_{\rm ph}\sim\frac{L_{\rm IR}}{4 \pi R_{\rm IR}^2c\epsilon_{\rm ph}}\simeq
10^{13}\left[\frac{10^{13}\mbox{ cm}}{R_{\rm IR}}\right]^2 \
\mbox{ cm}^{-3} \ ,
\end{equation}
for effective  collisions with protons.  

Protons interact with ambient photons also through the pair production
(Bethe-Heitler) process.  Although the cross-section of pair production is
larger than  the photo-meson cross section  by two orders of magnitude,
only a small, $10^{-3}$ fraction of the proton energy per interaction is
converted into electromagnetic secondaries. Therefore, at energies above
the photo-meson production threshold, hadronic interactions dominate
over the pair production.
The mean free path of protons through the photon field of density
given by Eq.(\ref{backgr}) is estimated
\begin{equation}
\Lambda_{p\gamma}\sim \frac{1}{\sigma_{p \gamma}  f n_{\rm ph}} \simeq 
10^{15}\left[\frac{R_{\rm IR}}{10^{13}\mbox{ cm}}\right]^2\mbox{ cm}  \, 
\end{equation}
(on average $\sigma_{p \gamma} f \simeq 10^{-28}$ cm$^2$, where 
$\sigma_{p \gamma}$ is the cross-section and $f$ is the inelasticity coefficient).

Since $\Lambda_{p\gamma}$ exceeds by two orders of magnitude the
linear size of the IR  source, $R_{\rm mm} \sim 10^{13} \ \rm cm$,
only 1 percent of the energy of protons is  converted into
secondary particles (at such high energies protons cannot be
effectively confined and therefore almost freely escape the IR  source).
Thus, in order to provide $\gamma$-ray luminosity at the level of
$L_\gamma\simeq 10^{35}$ erg/s, one has to require an injection power
of high energy protons
\begin{equation}
\label{p_flux}
L_p\sim \frac{\Lambda_{p\gamma}}{R_{\rm mm}} L_\gamma \approx
10^{37}\left[\frac{R_{\rm IR}}{10^{13}\mbox{ cm}}\right]\mbox{ erg/s}
\end{equation} 

\begin{figure}
\begin{center}
\resizebox{0.7\hsize}{!}{\includegraphics[angle=0]{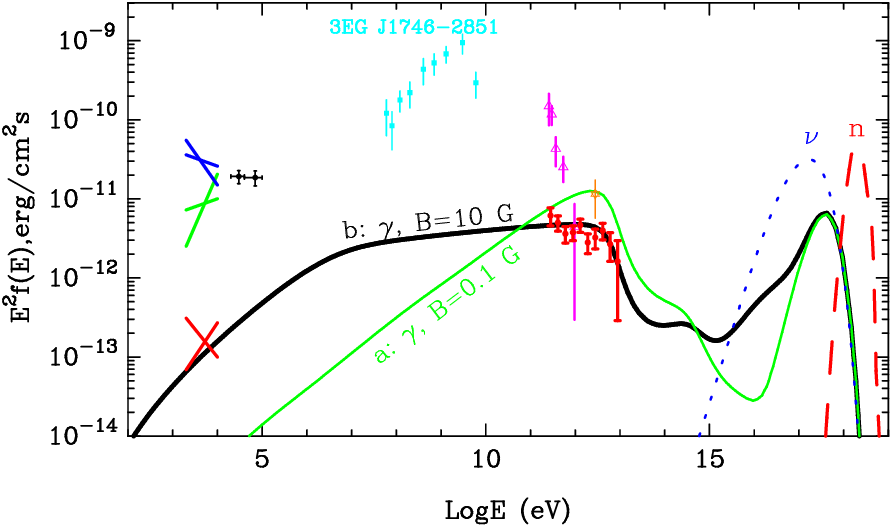}}
\caption{Broadband spectrum of $\gamma$-rays (solid lines), neutrons
(dash line) and neutrinos (dots) from Sgr A* due to interactions of
extremely high energy protons with ambient photon and magnetic fields.
Protons accelerated to energies $10^{18}$ eV in the regular magnetic
field close to the gravitational radius $R \sim R_{\rm g}$, propagate
through the infrared emission region of size $R=10 R_{\rm g}$. 
The calculations correspond to two assumptions for  the strength of the magnetic 
field in the region of the infrared emission:  $B=0.1$ G (a) and 
$B=10$ G (b).  The curves  represent  the  spectra
of cascade $\gamma$-rays emerging the source.  Note that because of
development of electromagnetic cascades, suppression of $\gamma$-rays
around $10^{14}$ eV is significantly less than one would expect from
simple photon-photon absorption effect.  The $\gamma$-ray and neutron fluxes are
corrected for the absorption of $\gamma$-rays due to interactions with
2.7 K CMBR (which results in the formation of a ``valley'' in the
spectrum at $10^{15}$ eV), and for the decay of neutrons, assuming 8 kpc
distance to the source. 
}
\label{fig:pgamma}
\end{center}
\end{figure}

For the parameters characterizing the IR  emission region, the
secondary $\pi$-mesons decay before interacting with the ambient
plasma, therefore their energy is immediately  released in the form of
neutrinos and $\gamma$-rays of energies $10^{17} - 10^{18} \rm eV$. 
The secondary neutrons are produced with somewhat larger energies.
While neutrinos and neutrons, as well as gamma-rays of energies below
$10^{12}$ eV and perhaps also above $10^{18}$ eV, escape freely the emission
region, gamma-rays of intermediate energies between $10^{12}$ and
$10^{18}$ eV, as well as secondary electrons from $\pi^\pm$-decays
effectively interact with the ambient photon and magnetic fields, and 
initiate  IC and/or (depending on the strength of the magnetic
field) synchrotron cascades.  The cascade
development stops when the typical energy of $\gamma$ rays is dragged
down to $\leq 10^{12}$ eV.  The energy spectra of gamma-rays 
calculated for two   different values of ambient magnetic field, $B=0.1$ and 10 G,  
are shown in Fig.~\ref{fig:pgamma}.

A distinct feature of this scenario is that TeV gamma-ray emission is
accompanied by detectable fluxes of ultrahigh energy neutrons, and possibly 
also  gamma-rays and neutrinos.  In particular, the luminosity of Sgr A*
in neutrons at $\geq 10^{18}$ eV can be as high as $L_{\rm n} \sim
10^{36}$ erg/s.  The corresponding   point-source flux 
of $10^{18}$ eV neutrons 
from  the direction of the Galactic Center, 
$F_{\rm n} \simeq 30$  neutrons/(km$^2$~yr),
exceeds by a factor of 100 the background of charged cosmic 
rays within 1 degree (the angular resolution of AUGER),
therefore   should be  detectable by AUGER   
as a  background-free signal.   
The  expected flux of $10^{17} - 10^{18}$ eV neutrinos from Sgr A* is also 
(marginally) detectable  with  AUGER  \cite{bertou2001}.

In this model the flux of  X-rays strongly depends
on the magnetic field in the region of the infrared source. For  the magnetic 
field $B \geq 10 \ \rm G$,  the secondary electrons are cooled effectively
which leads to the X/TeV energy flux ratio larger than 0.1.  
Therefore the interpretation of the TeV flux measured 
by H.E.S.S.   within this model predicts X-ray flux 
higher than the quiescent X-ray flux measured by Chandra.       

\subsubsection{Proton-proton scenario}
Acceleration of protons to extremely high energies, $E \sim 10^{18}$
eV, is a key element of the above scenario of proton-photon
interactions.  This implies existence of a strong magnetic field, $B
\geq 10^4$ G, in the compact region limited by a few gravitational
radii.  If the field close to the black hole is significantly weaker,
the efficiency of photo-meson processes is dramatically
reduced.  In this case interactions of protons with  protons and nuclei 
of ambient plasma become the main source of production of gamma-rays 
and electrons of  ``hadronic'' origin.

Protons can be accelerated to TeV energies also in the accretion disk,
e.g.  through strong shocks developed in the accretion flow. The
efficiency of gamma-ray production in this case is determined by the
ratio of accretion time $R/v_{\rm r} \sim 10^3 - 10^4$ s (depending on
the site(s) of particle acceleration and the accretion regime) to
the $p-p$ cooling time, 
\beq 
t_{\rm pp}=\frac{1}{\sigma_{pp} n c } \simeq
1.5 \times 10^{7}\left[\frac{10^8\mbox{ cm}^{-3}} {n}\right]\mbox{ s} \ , 
\eeq
where $n$ is the number density of the accretion plasma;  it depends
on the regime and geometry of accretion.
For any reasonable assumption concerning the density of the
ambient thermal plasma and the accretion regime, the efficiency of
conversion of energy of accelerated protons into secondary gamma-rays
and electrons is quite low, as small as  $10^{-4}$.  Therefore, even at most
favorable conditions, the acceleration rate of high energy protons should 
exceed $L_{\rm p} \approx 10^{39} \ \rm erg/s$ in order to provide
detectable fluxes of TeV gamma-rays.  Although the required
acceleration power is significantly larger than the total
electromagnetic luminosity of Sgr A*, yet it is still acceptable for 
a black hole of mass $\geq 10^6 M_\odot$.

\begin{figure}
\begin{center}
\resizebox{0.7\hsize}{!}{\includegraphics[angle=0]{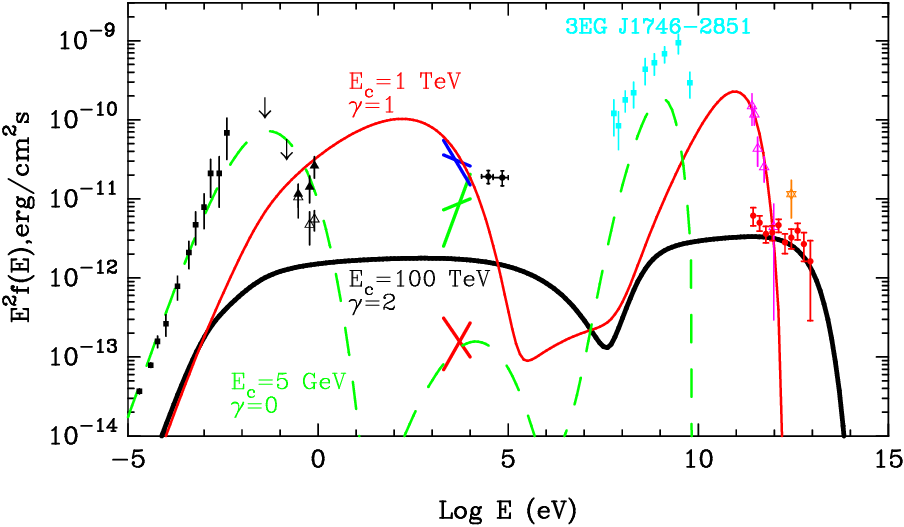}}
\caption{
Spectral energy distribution of the broad-band electromagnetic
radiation initiated by $p-p$ interactions in the accretion disk.  It is
assumed that the accelerated protons with spectrum $E^{-\Gamma}
\exp{(-E/E_c)}$ are injected into the thermal plasma of density $10^8
\ \rm cm^{-3}$, and together with the accretion flow cross the region
of the size $R\approx 10R_g$ cm and fall under the black hole
horizon after $10^4$ s.  The following parameters 
have been assumed:  (1) heavy solid curve: $\Gamma=2$, high energy
exponential cut-off at $E_c=100$ TeV, total acceleration rate $L_{\rm
p}=5 \times 10^{38} \ \rm erg/s$;  (2) thin solid curve:
$\Gamma=1$, $E_c=1$ TeV, $L_{\rm p}=10^{40} \ \rm
erg/s$; (3)  dashed curve: narrow ($\Gamma=0,E_c=5$~GeV) distribution of
protons, $L_{\rm p}=10^{40}$ erg/s. For all three cases  
the magnetic field is  assumed to be $B=10$ G. 
}
\label{fig:pp}
\end{center}
\end{figure}

The results of numerical calculation of the photon spectrum produced 
in $p-p$ interactions are shown in Fig. \ref{fig:pp}.
The shape of the overall spectral energy distribution, as well as
local spectral features depend both on the high energy
cutoff $E_0$ and the strength of the magnetic field.
If protons are accelerated to energies above 1 TeV then the
synchrotron radiation of secondary $e^+e^-$ pairs from $\pi$-meson
decays  extends  to hard X-ray domain.  In particular, acceleration of 
protons  beyond  $E_0 \geq  10 \ \rm TeV$ during a transient activity 
of the source  may result in a X-ray flare  with rather flat spectral energy distribution
(SED) like the flares  observed by XMM \cite{goldwurm2003}  and INTEGRAL
\cite{belanger2004} satellites (see Fig. \ref{fig:pp}).  Note
that although the characteristic radiative cooling time of protons
exceeds by many orders of magnitude the observed variability
timescales of X-rays,  $\Delta t \leq 1 \ \rm h$, the latter can be naturally
explained by the time when the accelerated protons confined in magnetic 
fields of the accretion flow cross the event  horizon.

If in the quiescent state the acceleration of protons is limited by
relatively low (GeV) energies, the maximum of the synchrotron
radiation moves towards IR and millimeter wavelengths. Interestingly,
if one assumes narrow, e.g.  Maxwellian type energy distribution of
protons, the resulting distribution of electrons also will be quite
narrow with mean energy approximately 10 times less than the proton
energy. Since the synchrotron cooling time of these electrons, 
\beq
t_{\rm synch}=4\times 10^4\left[\frac{10\mbox{ G}}{B}\right]^{3/2}
\left[\frac{10^{-2}\mbox{ eV}}{\epsilon}\right]^{1/2}\mbox{ s} \ ,
\eeq 
exceeds the typical dynamical timescale  of the
source, the radiative losses   do not deform significantly 
the production spectrum of secondary electrons, and therefore 
resulting synchrotron radiation at radio waves should have very hard
SED with spectral index $\approx 0.3$. Thus, 
with a certain combination of model parameters one 
can describe quite well the observed radio-to-IR spectrum by
synchrotron radiation of secondary electrons (see Fig. \ref{fig:pp}).  
This model is quite similar to  the traditional interpretation of the low-frequency 
radiation of Sgr A*  by directly accelerated electrons with Maxwellian type
distribution \cite{duschl1994}. The only difference is that in this case 
the narrow-energy distribution of electrons is resulting from  hadronic interactions 
of protons  with very flat  ($\Gamma < 1$) acceleration spectrum and 
cutoff  below 10 GeV.   While in the case of primary
electrons we do not expect significant gamma-ray emission (because both the IC
and bremsstrahlung channels of gamma-ray production are suppressed),
in the case of synchrotron radiation of secondary electrons one should expect 
very strong  $\pi^0$-decay gamma-ray emission at MeV/GeV energies (see
Fig. \ref{fig:pp}).
Interestingly,  the predicted MeV/GeV gamma-ray fluxes 
are quite close to the EGRET observations of the Galactic Center region. However, the
poor angular performance of these observations as well as lack of
information about the variability do not allow any certain conclusions
in this regard. On the other hand, the future observations by GLAST at
GeV energies should help to elucidate the origin of electrons
responsible for the radio to IR emission of Sgr A*.  An independent,
and to a certain extent more straightforward inspection of the
``hadronic'' origin of the broad-band SED can be provided by TeV
observations, and especially by future ``km3''  class neutrino detectors
(e.g. \cite{halzen})  which are expected to be  sufficiently sensitive for 
detection of a hard spectrum TeV neutrino signal 
(from decays of secondary $\pi^\pm$-mesons) 
with the flux comparable to the  TeV gamma-ray flux, i.e. 
$J_\nu( \geq 1$~TeV$) \sim 10^{-11} \ \rm \nu/cm^2 s$.   
A  clear observational signature of this scenario could  be 
robust correlation between TeV and X-ray radiation components  
- X-ray flares should be accompanied by TeV flares (unless the synchrotron cooling time significantly
exceeds the accretion time which can be realized if magnetic field is low and/or very fast accretion in the 
inner part of the disk).  In this regard, it is difficult
to overestimate the importance of continuous monitoring of of Sgr A*
at  gamma-ray energies  between 100 GeV to 10 TeV
by the H.E.S.S. telescope array and at  hard X-rays by Chandra, XMM and
INTEGRAL satellites with comparable sensitivities for detection of
flares at the energy flux level of $10^{-11} \ \rm erg/cm^2 s$ on hour
timescales. In particular, any detection of TeV gamma-rays with an energy flux exceeding 
the  X-ray flux by an order of magnitude, would be a strong evidence against the 
$p-p$ origin of TeV radiation.  

\subsection{Curvature Radiation - Inverse Compton  (CRIC) model} 

The  models of TeV gamma-ray emission associated with accelerated protons 
have a  drawback:  the efficiency of
conversion of proton power into electromagnetic radiation is rather
low. In the photo-hadron scenario the efficiency  is only 
0.1\% while in the proton-proton scenario it is even lower. The
radiative energy loss rate of electrons is much higher,  and therefore 
the models associated with accelerated electrons provide 
more economic ways  of  production of high energy gamma-rays. 
Obviously, these electrons should be accelerated to multi-TeV energies. 
This immediately constrains the strength of the chaotic component of the magnetic 
field in the  region of acceleration. Assuming that electrons are accelerated 
at a rate   $dE/dt\sim \kappa eB$ ($\kappa \leq 1$), 
from the balance of the acceleration and  synchrotron  energy loss rates one finds 
\begin{equation}
\label{max_e}
E_{\rm e} \le \frac{3^{3/4}m_e^2}{2^{3/4}e^{3/2}B^{1/2}}\simeq
1.5 \times 10^{13}\left[\frac{B}{10  \mbox{ G}}\right]^{-1/2} \kappa^{1/2} \mbox{ eV} \ .
\end{equation}
Thus, even in the case of maximum acceleration rate ($\kappa=1$) electrons cannot 
be accelerated to multi-TeV energies unless the random  B-field is less than 10 G.

The requirement of particle acceleration at the maximum rate imposes strong
restrictions on the possible acceleration mechanisms.  
In this regard,  acceleration in ordered electric and magnetic fields, 
e.g.  by the rotation-induced electric field near the black hole  provides 
maximum energy gain. Moreover, in the ordered  field  the 
energy dissipation  of electrons is   reduced to curvature radiation loses which 
increases the  maximum achievable energy of electrons to    
\beq
\label{e_curv}
E_e=\left[{3m_e^4 R^2 B \over 2e}\right]^{1/4}\simeq 10^{14}
\left[\frac{B}{10 \mbox{ G}}\right]^{1/4}\mbox{ eV}  \ .
\eeq
Note that  this estimate weakly depends on the strength of the magnetic field. 
On the other hand,  electrons may suffer also significant Compton losses
which would result in reduction of $E_e$ given by Eq.(\ref{e_curv}). Remarkably 
the radiative losses of both Curvature and Compton channels are released
in the form  of high energy and very high energy gamma-rays.
Indeed,  the inverse Compton scattering of 100 TeV electrons 
on infrared photons proceeds in the Klein-Nishina regime, and thus 
the IC spectrum  peaks at energy $E_\gamma \simeq E_e \simeq 100 \ \rm TeV$. 
At the same time the curvature radiation results in  
the  second peak in the spectrum which appears    at significantly lower  energies,   
\begin{equation}
\label{curv_e}
\epsilon_{\rm curv} ={3 E_e^3\over 2 m_e^3R}\simeq 2\times 10^8\left[\frac{E_e}{10^{14}\mbox{
eV}}\right]^3  \mbox{ eV} \ .
\end{equation} 
Below we will call the  scenario of production of Curvature and IC photons 
by electrons accelerated in regular magnetic/electric fields as {\em CRIC}  model.
Quantitative calculations of high energy radiation within the framework of this model
requires   ``self-consistent''  approach 
in which the spectrum of radiation is calculated simultaneously with the 
spectrum of  parent  electrons, because 
the spectrum of high-energy electrons  itself is determined by the balance
of acceleration and radiative  energy loss rates. 

\begin{figure}
\begin{center}
\resizebox{0.7\hsize}{!}{\includegraphics[angle=0]{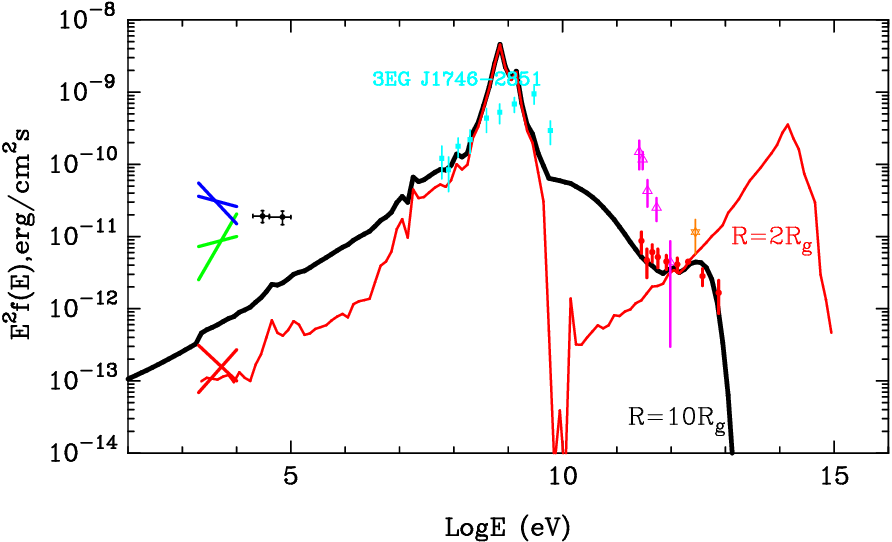}}
\caption{Broad-band spectral energy distribution of radiation
 produced by electrons within the {\em CRIC}  model (see the text). 
Thins solid  curve  -  gamma-ray production spectrum 
formed as superposition of the  Curvature and inverse Compton emission 
components which accompany  electron acceleration by the 
rotation-induced electric field within   ($R=2 R_{\rm g}$); 
heavy curve - the spectrum of 
gamma-rays modified after the passage of the infrared source  of size 
$R=10 R_{\rm g}$. The strength of the regular magnetic  field in the
electron acceleration  region is assumed  $B=10$ G.
The strength of  the random magnetic field in the region of infrared emission 
is assumed $B=30$ G. 
} 
\label{fig:sic}
\end{center}
\end{figure}

An example of such self-consistent computation is shown in Fig. 
\ref{fig:sic}.  The  calculations are performed  within the model
in which electrons are accelerated by the electric field induced by the 
black hole rotation (for details of the model see  Neronov et al. 2004). 
The  propagation of electrons in external 
electromagnetic field is calculated numerically taking into account the radiation reaction force.
The spectra of  synchrotron/curvature and inverse Compton radiation components 
are calculated at each point of the electron trajectory. The spectra shown in 
Fig. \ref{fig:sic} are result of summing up the spectra from  
about $10^{4}$ electron trajectories close (within $2$ gravitational radii) 
to the black hole horizon and subsequent propagation of the secondary gamma-rays 
through the infrared emission source  which is assumed to be confined   
within  $\sim 10$ gravitational radii. We assume 10 G regular
B-field in the acceleration zone, and 30 G chaotic field in the infrared source.  

In Fig.~\ref{fig:sic}  one  can see two distinct 
components in the gamma-ray {\em production} spectrum (thin solid line)  which 
sharply peak   at  $\sim 1$ GeV and 100 TeV.  
The GeV peak is due to the Curvature radiation, and 100 TeV peak  
is formed due to inverse Compton scattering that proceeds in the 
Klein-Nishina limit.   However the highest energy gamma-rays, 
$E_\gamma  \sim 10^{14}-10^{15}$ eV,  can not freely escape the source.
They effectively interact with infrared photons with production of 
 electron-positron pairs. The synchrotron radiation of these electrons in 
an irregular  field leads to the re-distribution of the initial gamma-ray spectrum  
(heavy solid line in Fig. \ref{fig:sic}). 

It should be  noted  that the fluxes shown in  Fig. \ref{fig:sic} are obtained under assumption 
of isotropic emission of electrons in the  acceleration zone. However,  
both  Curvature and inverse Compton radiation components produced during the
acceleration of electrons  are  emitted anisotropically.
This means that the presence or absence of a sharp feature in the spectrum
at GeV energies (see Fig. \ref{fig:sic}) depends on the configuration 
of the magnetic field in the acceleration zone and the viewing angle. 
The dependence of the inverse Compton component 
on the geometry of the source is less dramatic because the most of the 
energy of this component 
is absorbed  and redistributed in the  infrared source.  
The quantitative analysis of this effect 
is beyond the framework of this paper,  and will be discussed  elsewhere.

\section{Summary and Conclusions.}

The origin of  the TeV gamma-ray emission reported recently from 
the direction of the Galactic Center by three independent groups is 
not yet established.  Despite 
localization of the region of gamma-ray emission by H.E.S.S. -- within 3 arcmin for an extended 
source or  for a multiple-source cluster, and 1 arcmin for a point-like source --  
several objects remain as likely candidates for TeV emission. These are, in particular,
the central 10 pc region filled by dense gas  clouds and cosmic rays, the 
young supernova remnant  Sgr A East, the Dark Matter Halo, and the central 
compact radio source  Sgr~A*.  Although any of these sources may
contribute non-negligibly into the observed gamma-ray flux, in this paper 
we discuss a few possible TeV gamma-ray production scenarios related to Sgr~A*, 
namely in the immediate vicinity of the associated supermassive black hole. 

\begin{figure}
\begin{center}
\resizebox{0.7\hsize}{!}{\includegraphics[angle=0]{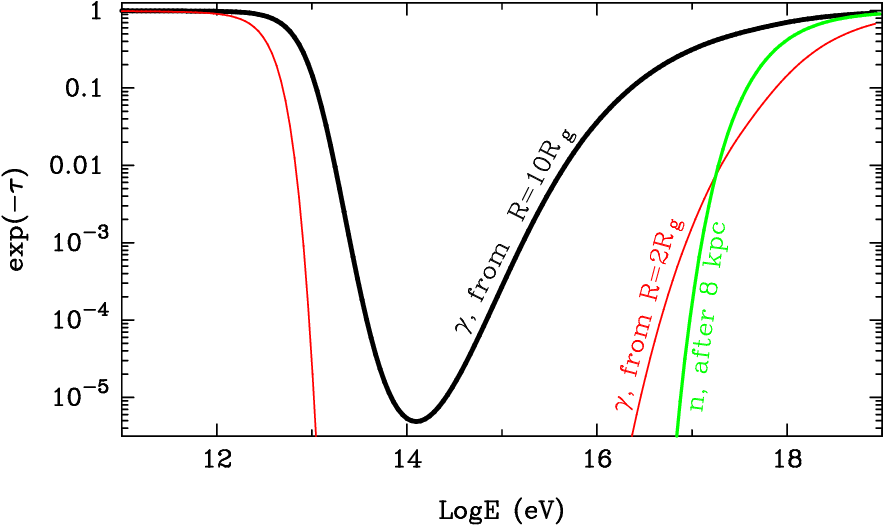}}
\caption{Attenuation of $\gamma$-rays in Sgr A* due to internal
photon-photon pair production dominated by interactions of high energy
gamma-rays with radiation of the compact infrared  source. 
The main uncertainty of the
optical depth $\tau$  is caused by  the uncertainty of
location of the infrared source.  Two solid curves marked ``$\gamma$'' are
calculated assuming that the infrared emission of Sgr A* is produced
within $10R_g$ and $2R_g$ around the central  black hole of mass 
$3 \times 10^6 M_\odot$.
External absorption of $\gamma$-rays due to interactions with the 2.7~K 
CMBR (not shown in the figure) is noticeable ($\sim e^{-1} \approx 1/3$) only at energies
around $10^{15} \ \rm eV$ \cite{gould1967}. 
However at these energies gamma-radiation is already strongly  suppressed 
due to the internal absorption.  The curve marked ``n'' shows attenuation of
the neutron flux $\exp{(-d/\Lambda)}$, where $\Lambda \approx 10
(E/10^{18}$ eV$) \ \rm kpc$ is the decay mean free path of a neutron of energy $E$, 
and $d=8 \rm kpc$ is the distance to the Galactic Center.}
\label{fig:opt_depth}
\end{center}
\end{figure}

\begin{figure}
\begin{center}
\resizebox{0.7\hsize}{!}{\includegraphics[angle=0]{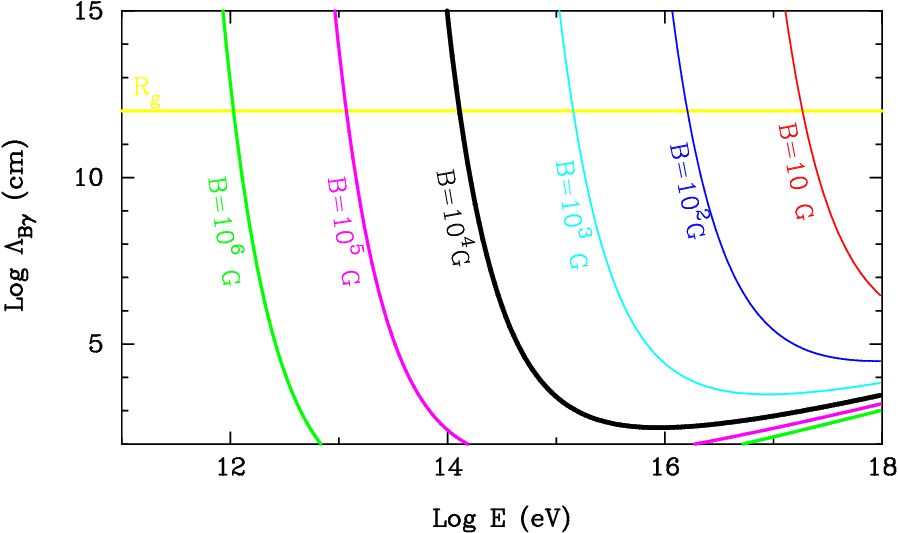}}
\caption{Absorption lengths  of $\gamma$-rays  
due to pair production  in magnetic field calculated 
for several  values of the strength of magnetic field (indicated at the curves). 
Horizontal line shows the gravitational radius of a $3\times 10^6M_\odot$ 
black hole.}
\label{fig:lambdaB}
\end{center}
\end{figure}

Production of high energy gamma-rays within 10 gravitational radii of a  black hole
(of any mass) could be copious  due to effective acceleration of particles by the 
rotation induced electric fields close to the event horizon or by strong shocks 
in the inner parts of the accretion disk. However, generally these energetic 
gamma-rays cannot escape the source because of severe absorption due to 
interactions with the dense low-frequency radiation through photon-photon pair 
production. This is true for both stellar mass and supermassive black holes. 
But, fortunately, the supermassive black hole in our Galaxy is an exception 
because of its unusually low bolometric luminosity. As shown in 
Sec. 2,  gamma-rays up to several  TeV can escape the source even if they are 
produced within a few gravitational radii (see Fig. \ref{fig:opt_depth}); the propagation 
effects related to the possible  cascading  in the photon field may extend 
the high energy limit to 10 TeV or even beyond. On the other hand  TeV gamma-rays are not  
absorbed in the magnetic field unless
the strength of the B-field in this region does  not exceed $10^5$ G
(see  Fig. \ref{fig:lambdaB}).
  
Thus, the   identification of the TeV signal  (or a fraction  of this signal)  
detected from the direction of the Galactic Center with 
Sgr~A*   would provide a  unique opportunity to study the high energy processes 
of particle  acceleration and radiation in the immediate vicinity of BH. 
The  transparentness  of Sgr A* for gamma-rays, as well as the recent reports of 
detection of TeV radiation from the direction of the Galactic Center, initiated 
the present  work with the main  objective  to explore possible processes of high energy 
gamma-ray production  within several  gravitational radii  of BH, and to study 
the impact of these processes on the formation of broad-band spectral energy 
distribution of  Sgr~A*.  We found that 
at least 3  scenarios can provide detectable TeV gamma-ray emission. 

(i) The first scenario is related to  protons accelerated 
to  $\sim 10^{18} \ \rm eV$. These protons can produce very high energy 
gamma-rays through Synchrotron and Curvature radiation. But in both cases 
the energy of gamma-rays does not extend to TeV energies. In the case of
synchrotron radiation it is limited by the so-called ``self-regulated'' cutoff around 300 GeV. 
On the other hand,  curvature radiation of TeV gamma-rays  in the black hole of mass 
$\sim 3 \times 10^6 M_\odot$ requires magnetic field exceeding $10^5$ G. But such
a strong field would prevent escape of gamma-rays due to electron-positron pair production.
A more effective channel for production of TeV gamma-rays is the photo-meson 
processes.  The mm-IR source in Sgr~A* is rather faint, but it is very compact, and 
therefore provides  sufficient density of seed photons for interactions with 
$10^{18} \ \rm eV$ protons. The efficiency of this process is not very high --
it does not exceed  0.1 per cent --  but the required power in accelerated 
protons of about  $10^{38} \ \rm erg/s$ is well below  the Eddington luminosity 
of the black hole.  This scenario predicts detectable fluxes of $10^{18} \ \rm eV$ 
neutrons, and perhaps also gamma-rays and neutrinos.

(ii) TeV gamma-rays can be produced  also by significantly lower energy protons,
accelerated by the electric field close to the gravitational radius  or by strong 
shocks in the  accretion disk. In this case the gamma-ray 
production  is dominated  by interactions of $\geq 10^{13} \ \rm eV$ 
protons with the accretion plasma. Because of the low efficiency of this 
process ($pp$ cooling time  is much longer than the characteristic dynamical time of 
the accretion glow),  the interpretation of the observed TeV fluxes
by  $\pi^0$-decay gamma-rays requires proton acceleration power as large as 
$10^{39} \ \rm erg/s$, which however is still below, 
at least by 4 orders  of magnitude,  the Eddington luminosity of the central black hole. 
This scenario predicts unavoidable TeV neutrino flux which can be detected by future 
neutrino detector NEMO located in the Northern Hemisphere.  This scenario 
predicts strong TeV-X-ray-IR  correlations. Therefore simultaneous observations of Sgr~A*
with  IR, X-ray and TeV telescopes with a goal of detection of 
sub-hour flares  at these three  energy bands, may provide evidence 
in favor or against this scenario.

(iii) Although interactions of electrons with ambient photon and magnetic fields 
proceed with much higher efficiency than the gamma-ray production by protons 
in the above two channels, a detectable TeV gamma-ray emission requires effective acceleration 
of electrons to energies  well above 1 TeV. A viable  site of  acceleration of  such 
energetic electrons by could be compact regions within a few gravitational radii, 
provided that electrons move along the lines of regular  magnetic field. 
In this case the electrons produce not only Curvature radiation which 
peaks around 1 GeV, but also inverse Compton gamma-rays  
(produced in the Klein-Nishina regime)  with the peak emission   around 100 TeV.  
However,  these energetic gamma-rays cannot escape the source.
They effectively interact with the infrared photons  and perhaps also with the 
magnetic field, produce   relativistic  electron and positron pair, and thus 
initiate electromagnetic cascades  inside the infrared source. The observed TeV gamma-rays
can be readily accommodated by this model  
from the point of view of both required acceleration power of electrons 
($\dot{W}_e \sim L_{\rm TeV} \sim 10^{35} - 10^{36} \rm erg/s$ and the reproduction of the 
observed spectral shape of  TeV gamma-rays.  Obviously, no neutrinos are expected 
within the framework of this model. 

Finally, we want to emphasize that  as long as the source(s)  of the TeV emission 
arriving from the direction of the Galactic Center is (are) not identified, 
the results  of this paper should not be treated  as an attempt to interpret the TeV data, 
but rather  should be  regarded as a  demonstration that the central supermassive 
black hole in our Galaxy is able indeed  to produce detectable fluxes of TeV gamma-rays.

%

\end{document}